\documentclass[epj]{svjour}

\usepackage{amsfonts}
\usepackage{amsmath,epsfig,bm}
\usepackage{graphicx}
\usepackage{color}

\usepackage[setpagesize=false]{hyperref}

\usepackage{tabularx} 
\newcolumntype{s}{>{\centering\arraybackslash\hsize=.666\hsize}X} 
\newcolumntype{b}{>{\centering\arraybackslash\hsize=1.333\hsize}X}

\usepackage{epsfig}

\newcommand{\be}{\begin{equation}}
\newcommand{\ee}{\end{equation}}
\newcommand{\ba}{\begin{eqnarray}}
\newcommand{\ea}{\end{eqnarray}}
\newcommand{\bd}{\begin{displaymath}}
\newcommand{\ed}{\end{displaymath}}

\begin{document}

\title{Net Baryon Fluctuations from a Crossover Equation of State}

\author{J. Kapusta\inst{1} \and  M. Albright\inst{1}\thanks{\emph{Present address:} Honeywell International, Golden Valley, MN 55422, USA} \and C. Young\inst{2}}
\institute{School of Physics \& Astronomy, University of Minnesota, Minneapolis, MN 55455,USA \and 
National Superconducting Cyclotron Laboratory, Michigan State University, East Lansing, MI 48824, USA}

\titlerunning{}
\authorrunning{}

\date{}

\abstract{We have constructed an equation of state which smoothly interpolates between an excluded volume hadron resonance gas at low energy density to a plasma of quarks and gluons at high energy density.  This crossover equation of state agrees very well with lattice calculations at both zero and nonzero baryon chemical potential.  We use it to compute the variance, skewness, and kurtosis of fluctuations of baryon number, and compare to measurements of proton number fluctuations in central Au-Au collisions as measured by the STAR collaboration in a beam energy scan at the Relativistic Heavy Ion Collider.  The crossover equation of state can reproduce the data if the fluctuations are frozen out at temperatures well below than the average chemical freeze-out.
}

\maketitle

\parindent=20pt

\section{Introduction}
\label{Intro}

The main motivation for a scan of beam energies approximately in the range $4 < \sqrt{s_{NN}} < 40$ GeV for collisions of large nuclei is to search for a critical end point in the QCD phase diagram.  Various theoretical arguments and models suggest the existence of a curve in the plane of temperature $T$ {\it versus} baryon chemical potential $\mu$ representing a line of first-order phase transitions.  This curve terminates in a second-order phase transition at some $T_c$ and $\mu_c$.  There is no agreement on the numerical values.  More generally one would like to create matter in heavy-ion collisions with moderate temperatures and high baryon densities to study the type of matter relevant to supernovae, proto-neutron, and neutron stars.  The NA61/SHINE Collaboration has begun a program to study collision systems ranging from p+p to Pb+Pb with fixed targets and beam momenta from 13 to
158 GeV/nucleon \cite{SHINE1,SHINE2,SHINE3}. The Nuclotron-based Ion Collider fAcility (NICA) is well positioned to focus on the interesting range $4 < \sqrt{s_{NN}} < 11$ GeV in collider mode \cite{NICAcollider}, preceded by fixed target experiments with beam energies between 1 and 4.5 GeV/nucleon \cite{NICAfixed}.

Recently we have constructed an equation of state in the form $P(T,\mu)$ which smoothly transitions from a gas of hadrons, with and without excluded volume effects, to a gas of quarks and gluons whose interactions are treated perturbatively \cite{Albright2014,Albright2015}.  It has several parameters which were adjusted to reproduce the pressure and interaction measure, or trace anomaly, calculated with lattice QCD.  With no further free parameters the crossover equation of state represented the lattice results at $\mu$ = 400 MeV just as well.  Treating hadrons as extended objects is better at reproducing the lattice results than treating them as point particles.  These results are described in section \ref{EOS}.

The STAR collaboration at the Relativistic Heavy Ion Collider (RHIC) has measured moments of the net-proton (proton minus anti-proton) multiplicity distributions in Au-Au collisions \cite{STAR_BES}.  These moments have been proposed as potential observables for critical behavior \cite{3rdmoments,nongaussian,4thmoments}.  The measurements were performed at beam energies $\sqrt{s_{NN}}$ = 7.7, 11.5, 19.6, 27, 39, 62.4, and 200 GeV.  In section \ref{STAR} we compare the variance, skewness, and kurtosis from our crossover equation of state to the STAR measurements for the 5\% most central collisions.  Fukushima \cite{Fukushima} made a similar comparison using a point particle hadron resonance gas and a relativistic mean field model with vector and scalar interactions.  

In order to best represent the STAR data we find that the fluctuations are determined at temperatures well below the average chemical freeze-out. 

Our conclusions are  contained in section \ref{conclusion}.

\section{Equation of State}
\label{EOS}

In \cite{Albright2014} we constructed a pressure $P(T,\mu)$ which includes a hadronic piece $P_h$, a perturbative QCD piece $P_{qg}$, and a switching function $S$ that ranges from 0 at low energy density to 1 at high energy density.
\be
P(T,\mu) = S(T,\mu) P_{qg}(T,\mu)  + \left[1 - S(T,\mu)  \right] P_h(T,\mu)
\label{switch}
\ee
The hadronic piece consists of a resonance gas comprising all known hadrons composed of up, down and strange quarks as given in the Particle Data Tables \cite{pdg2012}.  In one case the hadrons were treated as point particles.  We also used two excluded volume models for the resonance gas. For model I, the assumption is that the volume excluded by a hadron is proportional to its energy $E$ with the constant of proportionality $\epsilon_0$ (dimension of energy per unit volume) being the same for all species.  
For model II, the assumption is that the volume excluded by a hadron is proportional to its mass $m$.  We refer to these three hadronic equations of state as pt, exI and exII. Quantum statistics are used for the hadronic piece of the equation of state.

For the perturbative QCD piece we use the latest calculation which is valid up to order $\alpha_s^3 \ln \alpha_s$.  This piece involves two parameters, both relating to the energy scale used in the renormalization group running coupling.

The switching function must approach zero at low temperatures and chemical potentials and approach one at high temperatures and chemical potentials.  Figure 1 shows the trace anomaly, or interaction measure, $(\epsilon - 3P)/T^4$.  The hadron resonance gas represents the lattice result very well up to a temperature of about 150 MeV and then greatly exceeds it.  This is due to an increasing number of massive hadronic states with increasing temperature.  The perturbative QCD result represents the lattice result very well down to a temperature of about 220 MeV.  It deviates at lower temperature because the renormalization group running coupling becomes large.  Between these two limiting contributions there is a cusp around 190 MeV, and that is the challenge.  The switching function must also be infinitely differentiable to avoid introducing first, second, or higher-order phase transitions.  The functional form was chosen to be
\ba
 S(T, \mu) &=& \exp\{-\theta(T, \, \mu)\} \nonumber \\  
\theta(T, \mu) &=& \left[ \left( \frac{T}{T_0}\right)^n +   \left(\frac{\mu}{3 \pi T_0} \right)^n  \right]^{-1} 
\ea
with integer $n$.   This function goes to zero faster than any power of $T$ as $T \rightarrow 0$ (when $\mu = 0$).  It has two parameters, $T_0$ and $n$. 

\begin{figure}
\begin{center}
\includegraphics[width=0.99\linewidth]{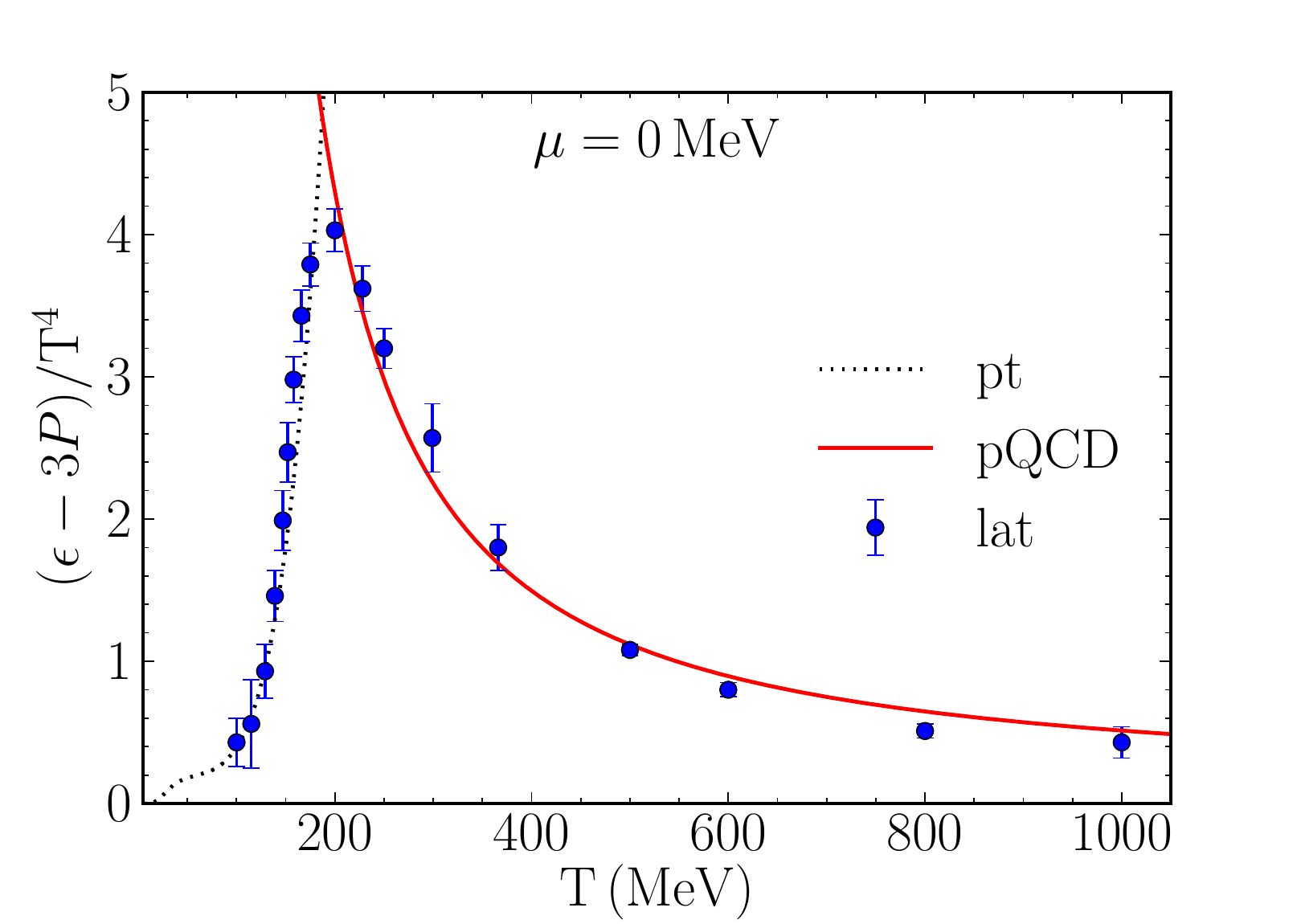}
\caption{ (color online) Trace anomaly normalized by $T^4$.  The dotted curve represents the parameter-free, point particle hadron resonance gas.  The solid curve represents perturbative QCD with 2 parameters adjusted to fit the lattice result taken from \cite{Borsanyi2010JHEP}. }
\end{center}
\label{anomalyHQ}
\end{figure}

We did a search on the parameters in each of the three models to obtain the best overall chi-square fit to both the pressure $P/T^4$ and the trace anomaly $(\epsilon - 3P)/T^4$ as functions of $T$ at $\mu = 0$ to the data in ref. \cite{Borsanyi2010JHEP}.  The point particle model best fits resulted in $n=4$ and $T_0=145.3$ MeV with a chi-squared per degree of freedom of 0.56.  The exI model best fit resulted in $n=5$, $T_0=177.1$ MeV, and $\epsilon_0^{1/4} = 306.5$ MeV with a chi-squared per degree of freedom of 0.34.  Nearly identical to that, the exII model best fit resulted in $n=5$, $T_0=177.6$ MeV, and $\epsilon_0^{1/4} = 279.7$ MeV with a chi-squared per degree of freedom of 0.34.  The value of $T_0$ for the point hadron gas is significantly smaller than for the excluded volume models.  The reason is that $P_h$ for the point hadron model grows much faster with $T$ than for the excluded volume models, and that fast growth must be cut-off by the switching function.  If an exponential mass spectrum of the Hagedorn type was used instead of only the known hadrons as presented in the Particle Data Tables, the $P_h$ for point hadrons would not even be defined above the Hagedorn temperature $T_H \approx 160$ MeV.  With the inclusion of excluded volumes for the hadrons this is not the case.  For that reason it is much better to treat hadrons as extended objects rather than point particles.  An interesting physical result of the excluded volume model fits is the resulting hard core radius of the proton and neutron.  For exI it is 0.580 fm while for exII it is 0.655 fm, very sensible numbers.

The comparison for the trace anomaly for $\mu = 0$ is shown in the top panel of figure 2.  With no free parameters the comparison for $\mu = 400$ MeV is shown in the bottom panel.  The visual fit of the excluded volume models is excellent; the fit is not as good for the point particle model.

\begin{figure}[ht]
\begin{center}
\includegraphics[width=0.99\linewidth]{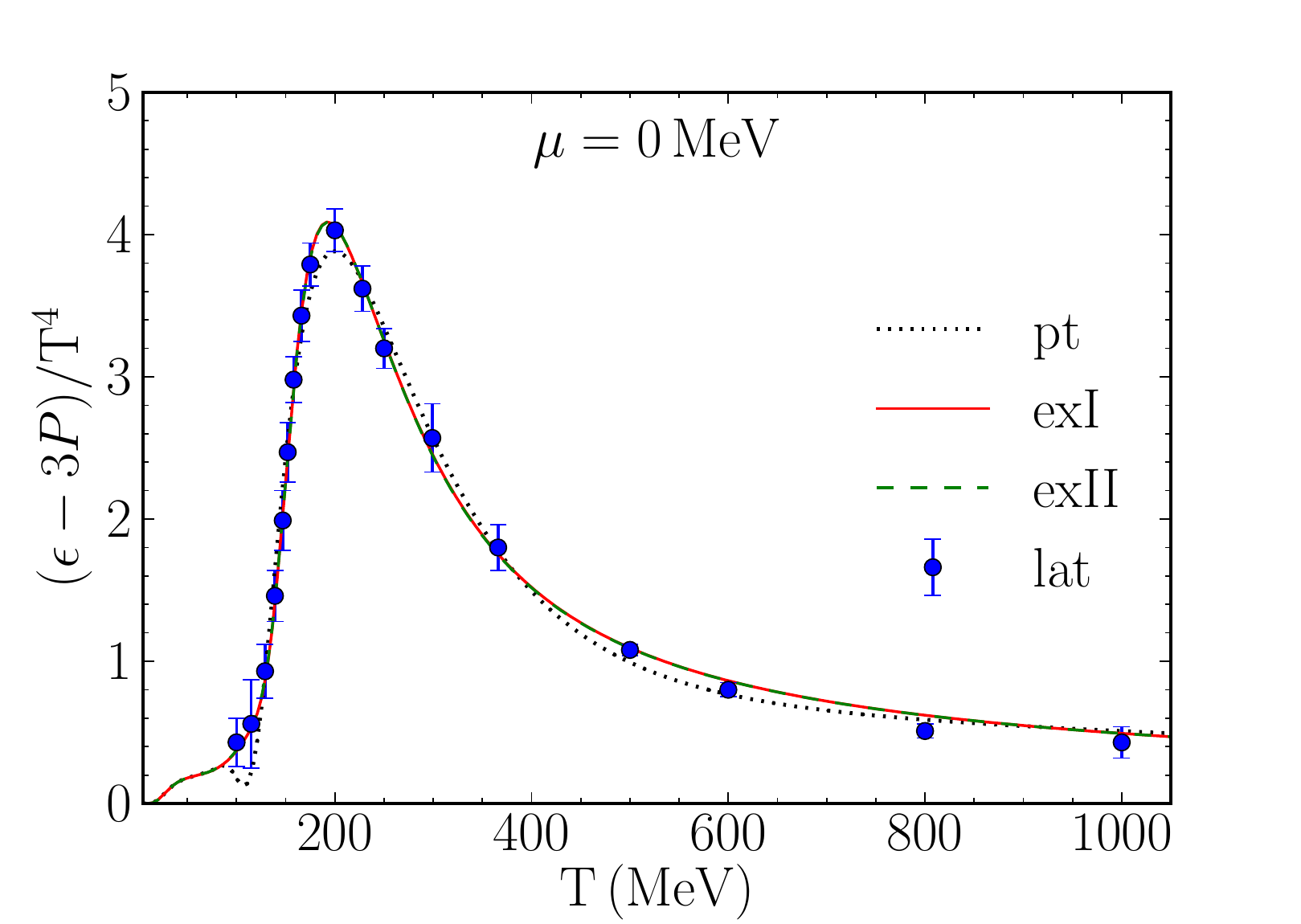}
\includegraphics[width=0.99\linewidth]{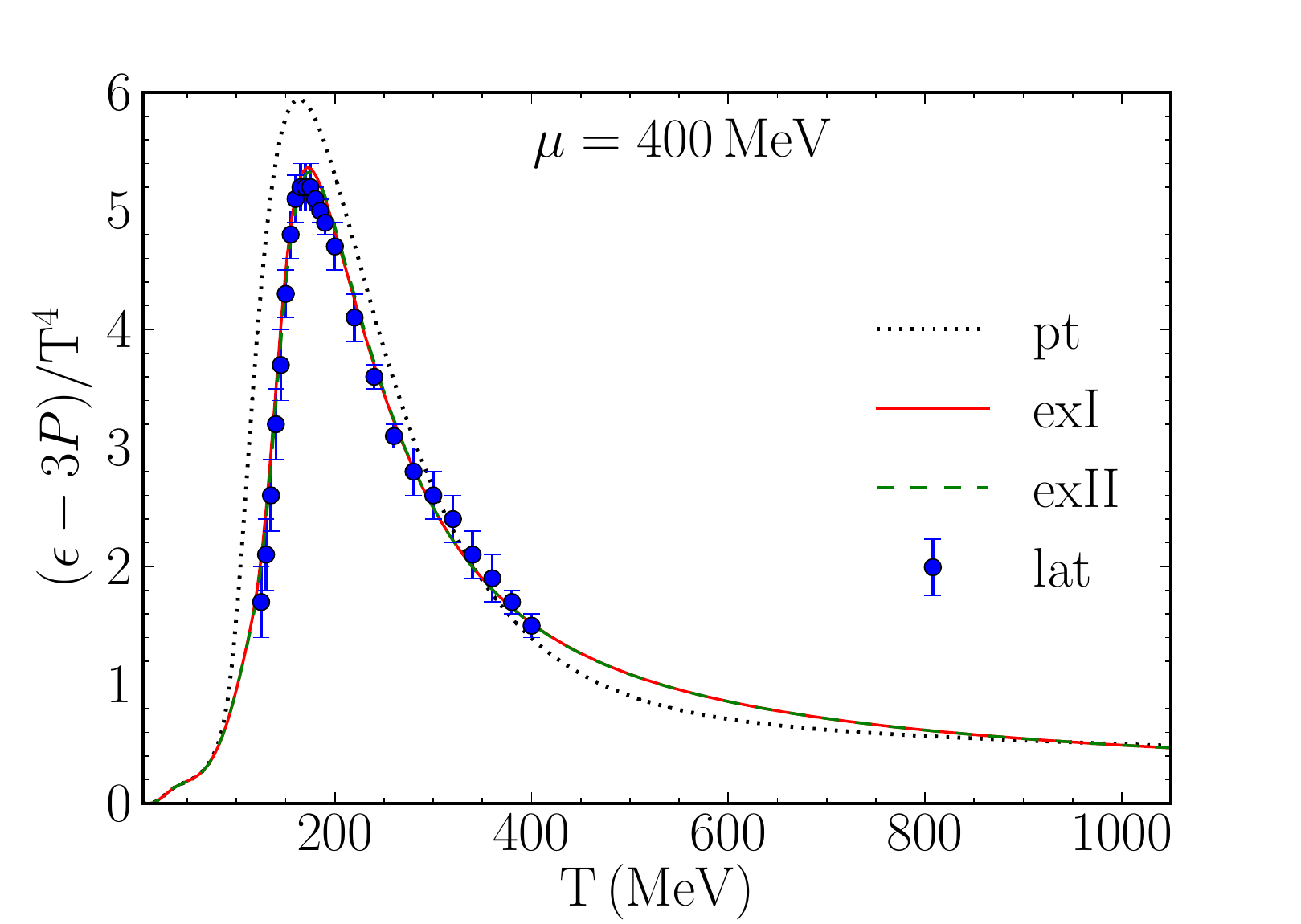}
\caption{ (color online) Trace anomaly normalized by $T^4$. The points are from lattice QCD \cite{Borsanyi2010JHEP} for $\mu = 0$ and from \cite{Borsanyi2012} for $\mu = 400$ MeV. }
\end{center}
\label{anomaly}
\end{figure}

A more sensitive test of the agreement between the crossover models and lattice QCD is the speed of sound given by $c_s^2 = \partial P/\partial \epsilon$, where the derivative is taken at constant entropy per baryon.  In terms of the susceptibilities
\be
\chi_{xy} = \frac{\partial^2 P(T,\mu)}{\partial x \partial y}
\ee
where $x$ and $y$ may be $T$ or $\mu$, it is
\be
c_s^2 = \frac{n_B^2 \chi_{TT} - 2 s n_B \chi_{\mu T} + s^2 \chi_{\mu\mu}}{w (\chi_{TT} \chi_{\mu\mu} - \chi^2_{\mu T})} \,.
\label{speed}
\ee
A comparison to lattice QCD is shown in figure 3 for $\mu=0$ (top panel) and for $\mu=400$ MeV (bottom panel).  The crossover equation of state using point hadrons has an unnatural temperature dependence around 115 MeV and so we do not show it here.  In both panels the dashed curve shows the sound speed for a massive non-interacting pion gas, which is the natural limit as the temperature goes to zero.  
\begin{figure}[thp]
\begin{center}
\includegraphics[width=0.99\linewidth]{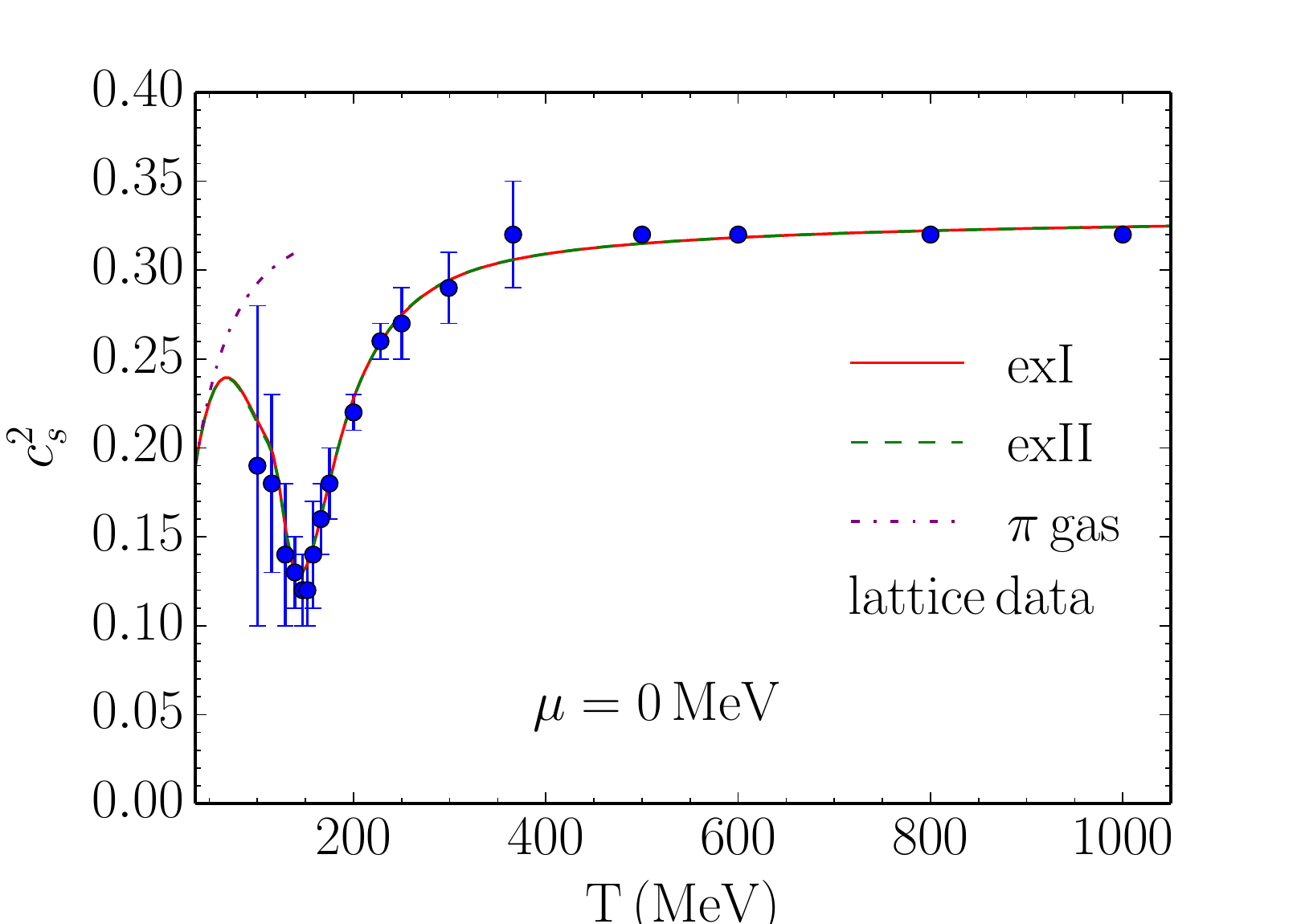}
\includegraphics[width=0.99\linewidth]{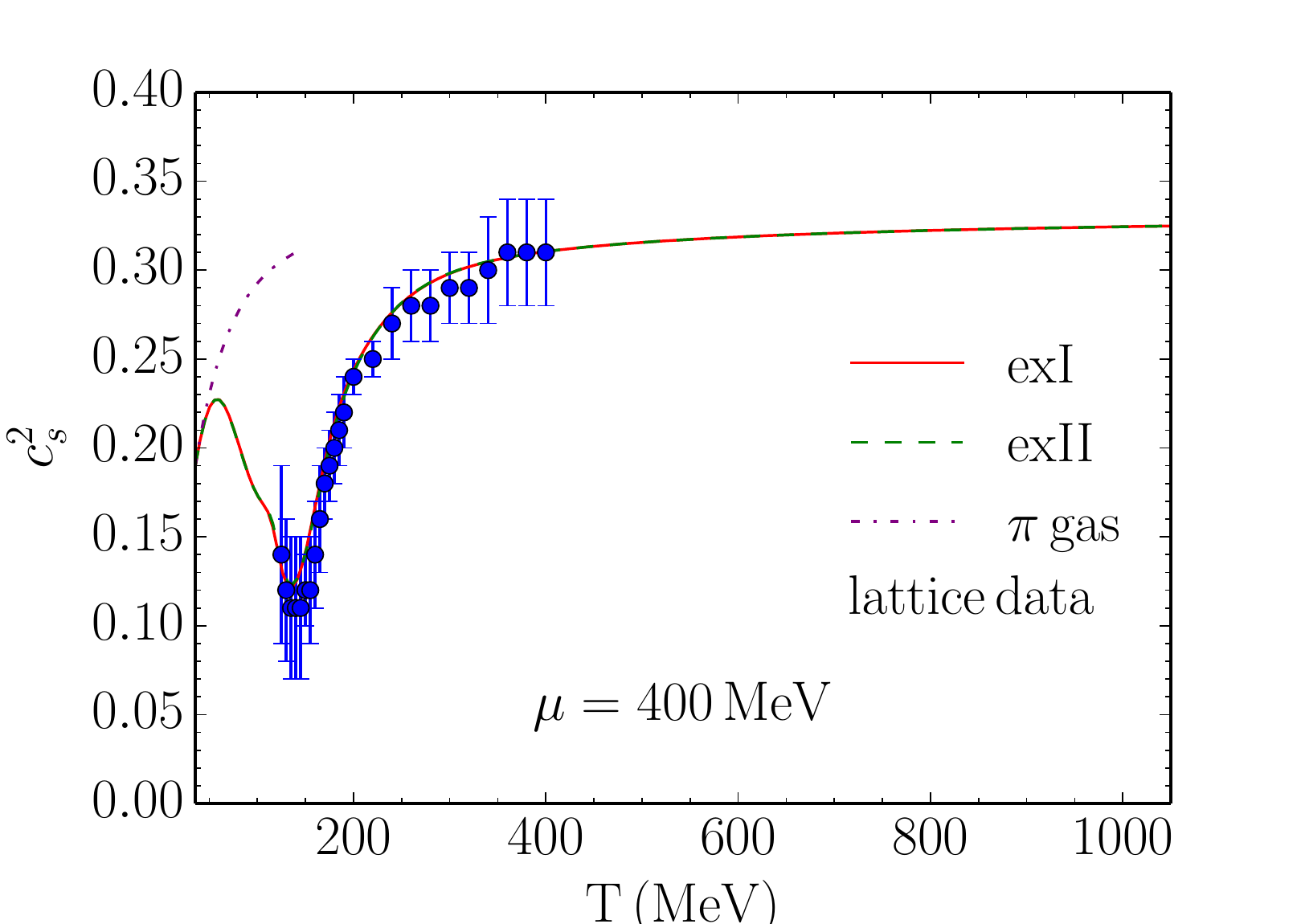}
\caption{(color online) The square of the sound speed as a function of temperature.  The points are from lattice QCD \cite{Borsanyi2010JHEP} for $\mu = 0$ and from \cite{Borsanyi2012} for $\mu = 400$ MeV.  The dashed line is for a noninteracting massive pion gas. }
\end{center}
\label{crossover_cs2}
\end{figure}

The mean $M$ and a variance $\sigma^2$ of the baryon number are
\be
M = V \chi_{\mu}
\ee
and 
\be
\sigma^2 = V T \chi_{\mu\mu}
\ee
where $V$ is the volume.  It is convenient to take ratios such that the volume cancels.  The next higher moments are the skewness and kurtosis.  For a finite size system, such as the matter formed in heavy-ion collisions, it is convenient and useful to consider the scaled skewness and kurtosis which are intensive quantities.  For a system in equilibrium they are
\be
S \sigma = T \frac{\chi_{\mu\mu\mu}}{\chi_{\mu\mu}}
\ee
and
\be
\kappa \sigma^2 = T^2 \frac{\chi_{\mu\mu\mu\mu}}{\chi_{\mu\mu}} \,.
\ee
One general statement that can be made is $S \sigma = 0$ if the chemical potential vanishes.  The reason is that the pressure is an even function of the chemical potential, hence all odd derivatives vanish when $\mu = 0$.

Consider a point particle hadron resonance gas.  When Boltzmann statistics are adequate for the baryons, the pressure takes the form
\be
P(T,\mu) = P_b(T) \cosh(\mu/T) + P_m(T)
\ee
in an obvious notation.  Then $S \sigma = \tanh(\mu/T)$ and $\kappa \sigma^2 = 1$.  This is a very strong prediction.  For comparison, consider a noninteracting gas of massless quarks and gluons.  The pressure takes the form
\be
P(T,\mu) = a_0 T^4 + a_2 T^2 \mu^2 + a_4 \mu^4
\ee
where $a_0$, $a_2$, and $a_4$ are constants.  Then
\be
S \sigma = 12 a_4 \mu T / (a_2 T^2 + 6 a_4 \mu^2)
\ee
\be
\kappa \sigma^2 = 12 a_4 T^2 / (a_2 T^2 + 6 a_4 \mu^2) \,.
\ee
The dependence on the ratio $\mu/T$ is very different for these two equations of state.  Comparison of the crossover predictions to the STAR data will be made in the next section.

\section{Comparison to STAR Data}
\label{STAR}

\begin{figure}[th]
\begin{center}
\includegraphics[width=0.99\linewidth]{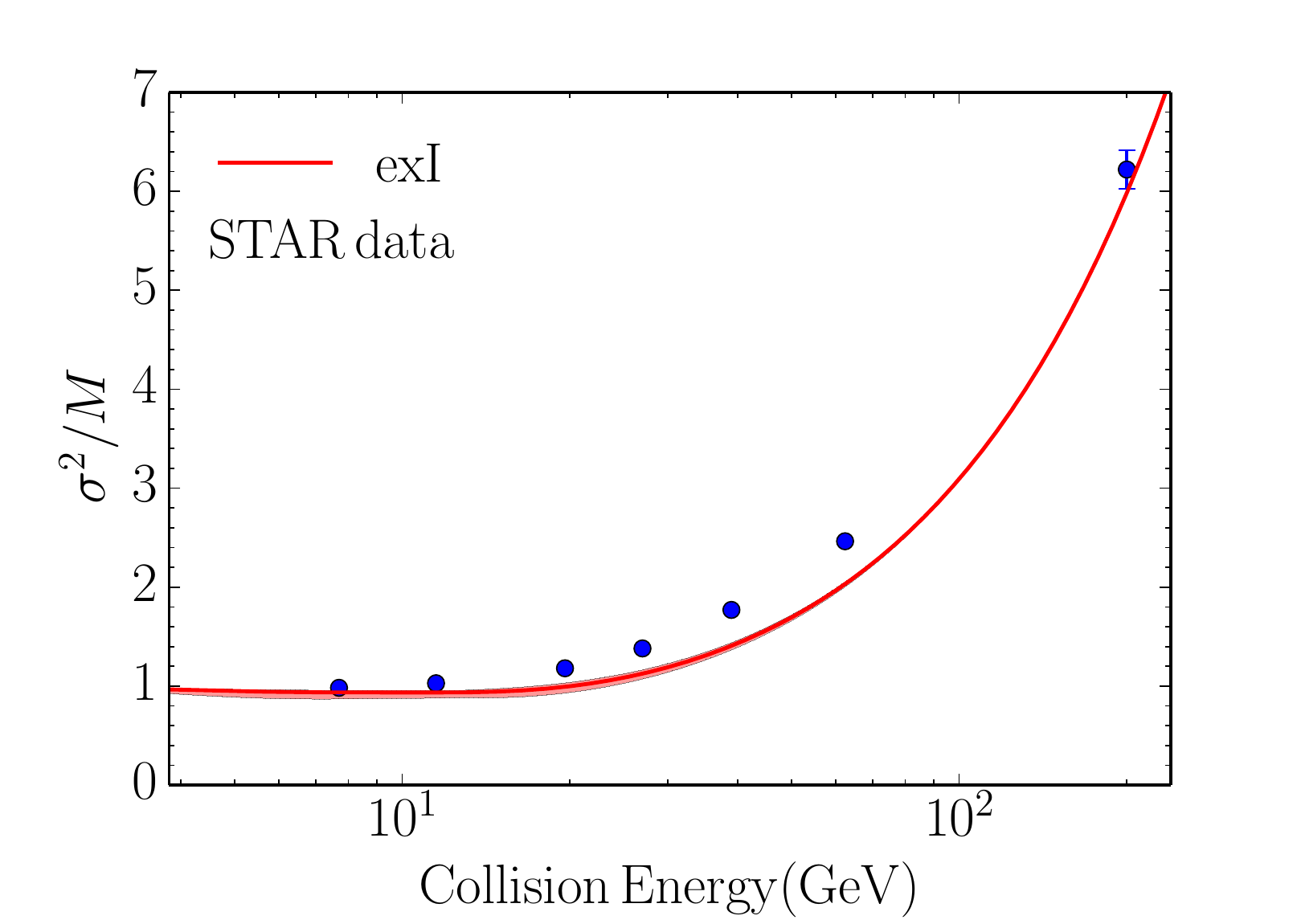}
\caption{(color online) Ratio of variance to mean for the crossover equation of state compared to the measurements by the STAR collaboration. The energy dependence of the temperature and chemical potential are determined as in Eq. (\ref{Tchem}) but with a temperature which is 26 MeV lower ($a = 140$ MeV).}
\end{center}
\label{variance_STAR_140}
\end{figure}
In this section we compare to the data taken during the first beam energy scan at RHIC by the STAR collaboration \cite{STAR_BES}.  In order to make the comparison we need to have an estimate of the temperature and chemical potential at the time the fluctuations are frozen out.  Following Fukushima \cite{Fukushima} we use the conditions at the time of average chemical freeze-out as presented in \cite{freezeoutcurve}.  The chemical potential is parameterized as a function of $\sqrt{s_{NN}}$ by
\be
\mu(\sqrt{s_{NN}}) = \frac{d}{1 + e \sqrt{s_{NN}}}
\ee
and then the temperature by
\be
T(\mu) = a - b \mu^2 - c \mu^4
\label{Tchem}
\ee
The five constants in these parameterizations are $a = 0.166$ GeV,  $b = 0.139$ GeV$^{-1}$, $c = 0.053$ GeV$^{-3}$, $d = 1.308$ GeV, and $e = 0.273$ GeV$^{-1}$.

It turns out that the measured skewness and kurtosis are not reproduced by the crossover equation of state with excluded volume corrections.  The data is larger than the theoretical predictions by about a factor of 2 at all beam energies.  Why is that?

One possibility is that the baryon fluctuations are not frozen out at the same time as average chemical freeze-out.  Indeed, it is well known that kinetic freeze-out occurs after average chemical freeze-out when the temperature is lower \cite{kinetic1,kinetic2,kinetic3}.  To get closer agreement with the STAR data, we ought to consider the possibility that the baryon fluctuations freeze-out occurs at lower temperatures than assumed earlier.

As an example, consider changing the parameter $a$ in Eq. \ref{Tchem} from 166 to 140 MeV.  Such a lower temperature is roughly consistent with kinetic freeze-out \cite{kinetic1,kinetic2,kinetic3}.  The results for the variance over the mean are shown in figure 4 for the crossover equation of state exI (the results for exII are nearly identical).  The agreement is acceptable.  

\begin{figure}[thp]
\begin{center}
\includegraphics[width=0.99\linewidth]{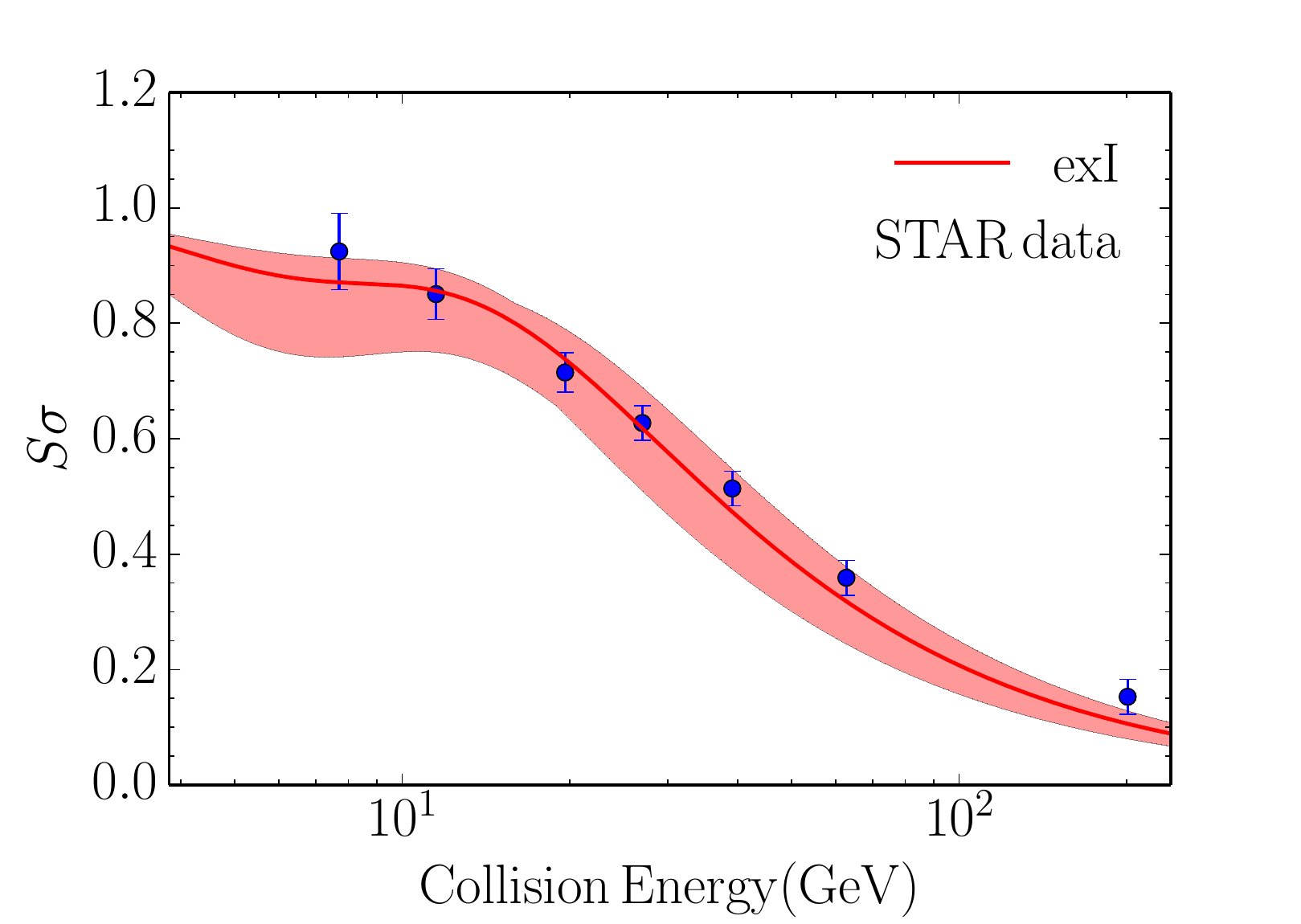}
\caption{(color online) The skewness for the crossover model exI.  The shaded region shows the uncertainty when fitting the crossover equation of state parameters to lattice QCD at zero chemical potential.  The energy dependence of the temperature and chemical potential are determined as in Eq. (\ref{Tchem}) but with a temperature which is 26 MeV lower  ($a = 140$ MeV).}
\end{center}
\label{crossover_skewness_STAR_140errorband}
\end{figure}

\begin{figure}[h]
\vspace{-0.062cm}
\begin{center}
\includegraphics[width=0.99\linewidth]{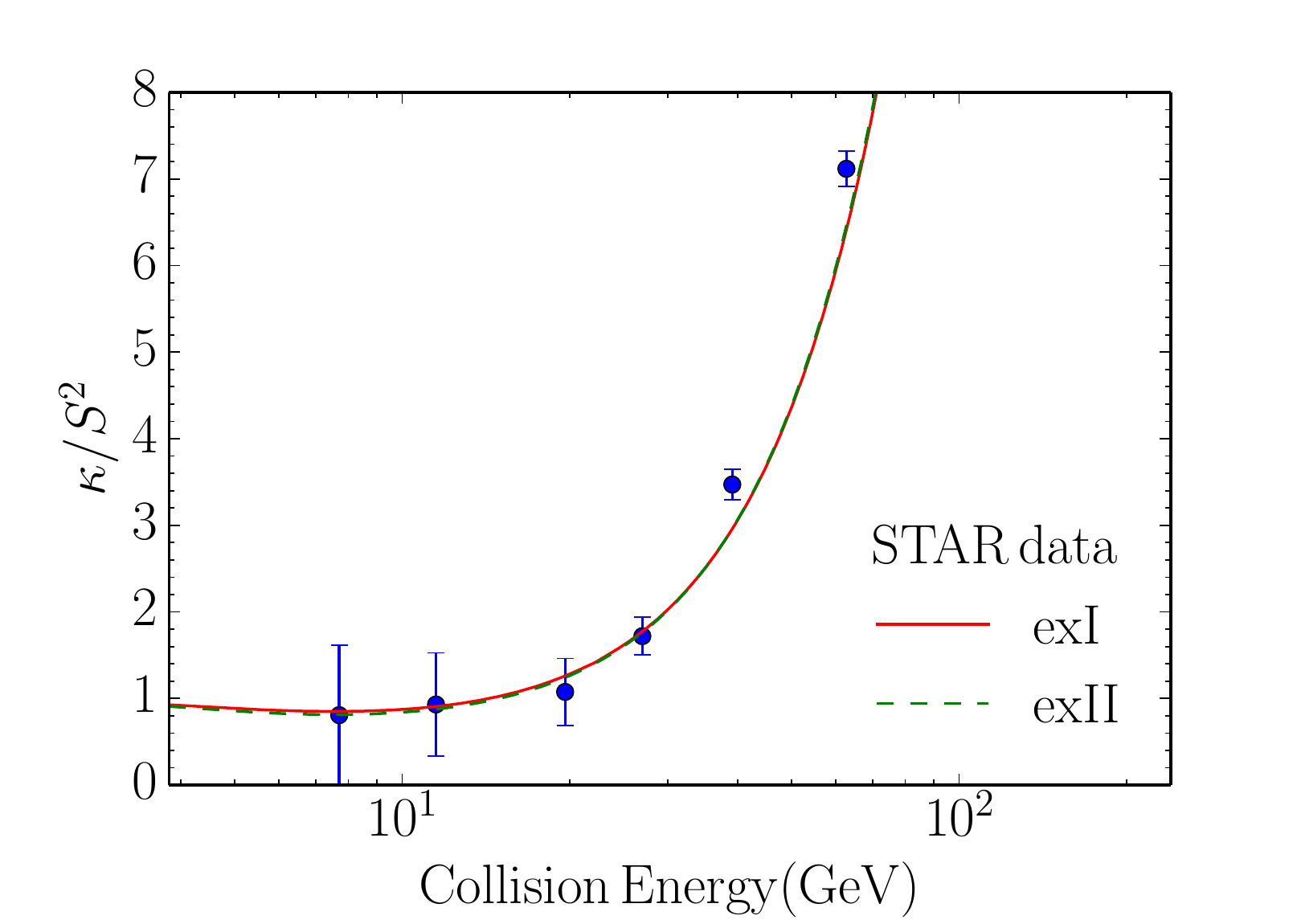}
\caption{(color online) Ratio of kurtosis to the square of skewness for the two crossover equations of state compared to the measurements by the STAR collaboration. }
\end{center}
\label{crossover_ratio_STAR_140}
\end{figure}
The skewness is plotted in figure 5, also with $a = 140$ MeV.  Once again, agreement is acceptable.

It is interesting to plot the ratio $\kappa/S^2$ since this is independent of the variance.  The result, using $a = 140$ MeV as before, is shown in figure 6.  The agreement at the four lowest beam energies is excellent.  The biggest discrepancies are at the highest beam energies.  As there is no discrepancy for $\sqrt{s_{NN}}$ = 7.7, 11.5, 19.6, and 27 GeV, it would be difficult to argue for a critical point in this energy range - at least under the assumptions made here.

Suppose that at each beam energy we wanted to fit the experimental measurements of $\kappa \sigma^2$ and $S \sigma$.  Assuming an equation of state, one can always find a $T$ and $\mu$ at each energy to fit this data.  Using the crossover equation of state exI, we show the results in figure 7.  The large uncertainty at large $\mu$ is a direct consequence of the experimental uncertainties.

\begin{figure}[t]
\begin{center}
\includegraphics[width=0.99\linewidth]{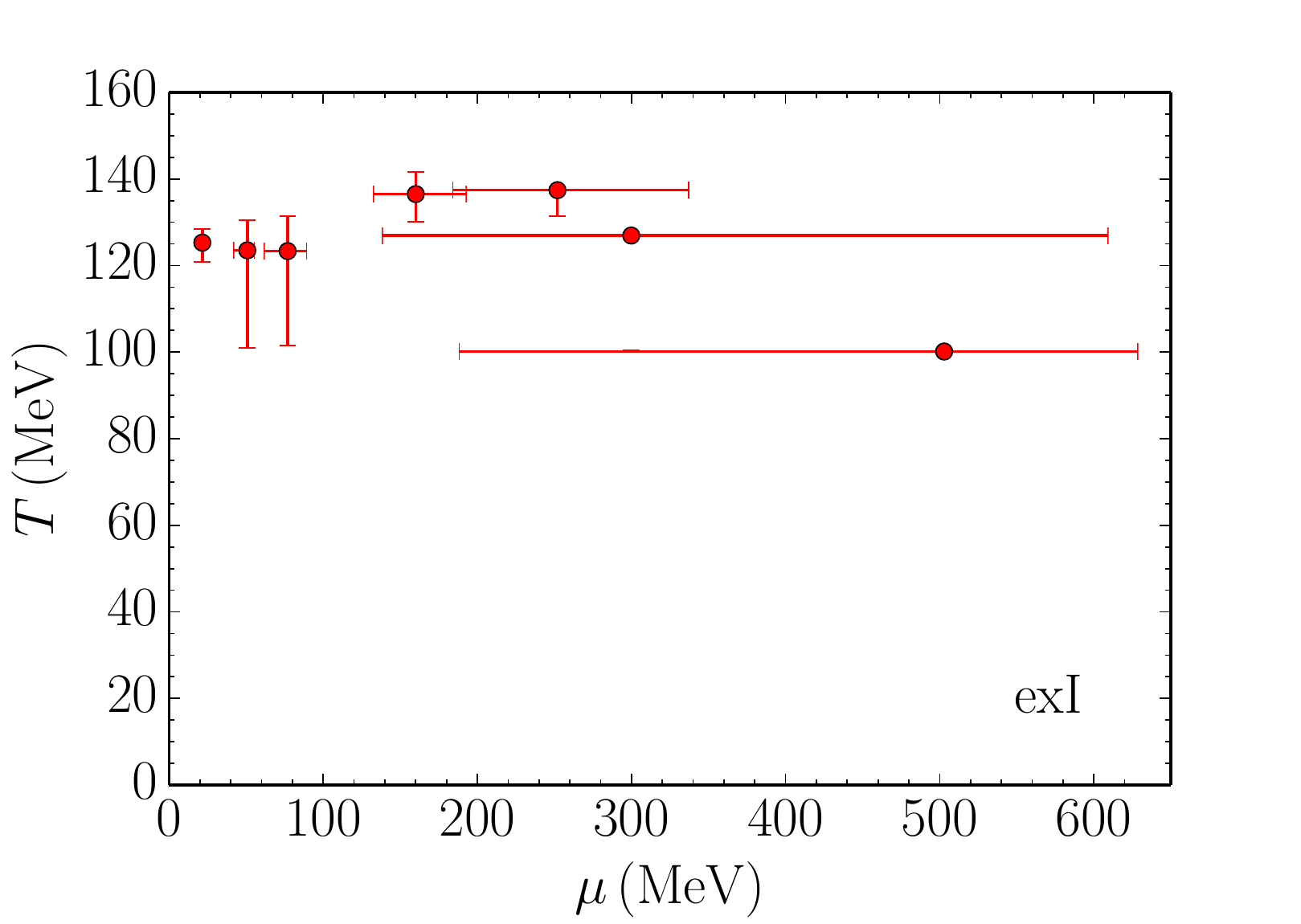}
\caption{(color online) A fit to the STAR measurements of the temperature and chemical potential at each beam energy using the crossover equation of state exI.}
\end{center}
\label{T_vs_mu_STAR}
\end{figure}

\section{Conclusion}
\label{conclusion}

We compared crossover equations of state with lattice QCD results and with measurements of the first beam energy scan at RHIC.  The previously constructed crossover equations of state interpolated between perturbative QCD at high energy density to a hadronic resonance gas at low energy density.  The hadronic resonance gas, with excluded volume effects included, gave excellent agreement with the sound speed as calculated on the lattice.  Less satisfactory results were found when hadrons were treated as point particles.

Skewness and kurtosis of the baryon number fluctuations are very sensitive measures of the equation of state because they involve third and fourth derivatives of the pressure with respect to the chemical potential.  The crossover equations of state are in quantitative agreement with experimental measurements of the skewness and kurtosis when it is assumed that the fluctuations are frozen out at lower temperatures.

There are obvious questions that deserve further investigation.  How accurate are the lattice QCD results, especially at nonzero chemical potential?  How accurately does the crossover equation of state need be known to replicate the lattice QCD equation of state, given that the skewness and kurtosis involve third and fourth order derivatives of the pressure?  Our study does not include the requirement that the system have zero net strangeness, which is probably not a major factor but still needs investigation.  A serious issue is the phenomenology connecting the experimental measurements to the equation of state.  For example, the experimental measurements have a lower momentum-space cutoff for protons of 400 MeV.  Such cutoff effects have been investigated in \cite{Garg2013} and \cite{Bhattacharyya2014}.  However, in general these cutoff effects are not so  straightforward when the equation of state includes interactions.

Our study does not suggest evidence for a critical point because it assumes a smooth crossover between  hadrons and quarks and gluons.  Clearly there is much work to be done, such as modifying the switching function $S$, as in \cite{ABMN2008}, to allow for a critical point to see what effect it would have on the variance, skewness, and kurtosis as functions of temperature and chemical potential.  Higher statistics and lower energies attainable at NICA are also required.  

\section*{Acknowledgements}

MA and JK are supported by the US Department of Energy (DOE) under Grant No. DE-FG02-87ER40328.  CY is supported by the US Department of Energy (DOE) under Grant No. DE-FG02-03ER41259.

\end{document}